\documentclass[pre,12pt]{revtex4-1}

\usepackage{graphicx}
\usepackage{latexsym}
\usepackage[centertags]{amsmath}
\usepackage{amsfonts}
\usepackage{amssymb}
\usepackage{amsthm}
\usepackage{newlfont}

\newcommand{\bu}{{\mathbf u}}
\newcommand{\bv}{{\mathbf v}}

\newcommand{\hatn}{\hat{n}}
\newcommand{\hatt}{\hat{t}}
\newcommand{\gat}{\gamma_a}
\newcommand{\dat}{\Delta}

\newcommand{\ri}{R_I}
\newcommand{\eps}{\epsilon}

\begin{document}

\title{Coulomb friction driving Brownian motors}

\date{\today}


\author{Alessandro Manacorda}
\affiliation{Dipartimento di Fisica,
Universit\`a ''Sapienza'', p.le A. Moro 2, 00185 Rome, Italy} 

\author{Andrea Puglisi}
\affiliation{Istituto dei Sistemi Complessi - CNR and Dipartimento di Fisica,
Universit\`a ''Sapienza'', p.le A. Moro 2, 00185 Rome, Italy} 
\affiliation{Kavli Institute for Theoretical Physics China, CAS, Beijing 100190, China} 

\author{Alessandro Sarracino}
\affiliation{Istituto dei Sistemi Complessi - CNR and Dipartimento di Fisica,
Universit\`a ''Sapienza'', p.le A. Moro 2, 00185 Rome, Italy} 
\affiliation{Kavli Institute for Theoretical Physics China, CAS, Beijing 100190, China} 
\affiliation{Laboratoire de Physique Th\'{e}orique et de la Mati\`{e}re Condens\'{e}e, CNRS UMR 7600,
case courrier 121, Universit\'e Paris 6, 4 Place Jussieu, 75255 Paris Cedex}

\begin{abstract}
We review a family of models recently introduced to describe
Brownian motors under the influence of Coulomb friction, or more
general non-linear friction laws. It is known that, if the heat bath
is modeled as the usual Langevin equation (linear viscosity plus white
noise), additional non-linear friction forces are not sufficient to
break detailed balance, i.e. cannot produce a motor effect. We discuss two
possibile mechanisms to elude this problem. A first possibility,
exploited in several models inspired to recent experiments, is to replace the heat
bath's white noise  by a ``collisional noise'', that is the
effect of random collisions with an external equilibrium gas of
particles. A second possibility is enlarging the phase space, e.g. by
adding an external potential which couples velocity to
position, as in a Klein-Kramers equation. In both cases, non-linear
friction becomes sufficient to achieve a non-equilibrium steady state
and, in the presence of an even small spatial asymmetry, a motor
effect is produced.
\end{abstract}


\maketitle

\section{Introduction}

Thermal fluctuations rule the dynamics of micro- and mesoscopic
objects. In equilibrium conditions, where detailed balance (DB) holds,
the effect of fluctuations can be described exploiting the underlying
symmetries for time-reversal and time-translation. In nonequilirbium
steady states, only the latter symmetry holds and a rich phenomenology
can take place, which would be ruled out in equilibrium conditions.
One of the main interesting behaviors, peculiar to nonequilibrium
dynamics, is the possibility of rectifying unbiased fluctuations, when
a spatial asymmetry is present in the system.

Among the many ways to break the time-reversal symmetry in statistical
models, the action of Coulomb friction has been recently put in
evidence~\cite{H05,dGen05,gnoli,gpt13}. This is a form of energy dissipation
that is observed in the relative motion of sliding surfaces of
macroscopic objects, and its microscopic theory is still at the core
of an intense debate~\cite{VMUZT13}.  Here we consider the macroscopic
modelization of the frictional force, namely we consider it as a
constant force opposite to the motion direction. The presence of
Coulomb friction introduces a strong nonlinearity in the system and is
a source of dissipation that can drive the system out of equilibrium.

In Section II we introduce our general model with thermal baths and
non-linear friction, discussing the conditions to break detailed
balance. In the same section we also put our model in the context of
previous models and experiments. In Section III we make a particular
choice for the heat bath, in the form of a dilute gas at equilibrium.
Three specific examples are discussed in detail. In Section IV the
role of heat bath is played by white noise with linear drag, in the
presence of an external potential which makes the system spatially
inhomogeneous. Conclusions are drawn in Section V.

\section{A general model with non-linear friction}

We consider a general model describing the motion of a probe, also
called ``tracer'' or ``intruder'', of mass $1$, in contact with a heat
bath and/or a gas of particles at equilibrium.  Beside the interaction
with the baths, the motion of the object may take place in a spatial
potential and is affected by some kind of non-linear dissipative
force. We assume that memory effects are negligible and the system is
a Markovian stochastic process.  The general differential equation
describing the probability density function of the system is therefore
\begin{subequations} \label{model}
\begin{align} 
\frac{\partial P(x,v,t)}{\partial t}&=-\frac{\partial}{\partial x}[vP(x,v,t)]-\frac{\partial}{\partial v}\{[F_{nl}(v)- U'(x)]P(x,v,t)\} +\\ & J_{Lang}[v|P(x,v,t)]+ J_{col}[v|P(x,v,t)] \nonumber \\
J_{Lang}[v|P(x,v,t)]&= -\frac{\partial}{\partial v}[(-\gamma v)P(x,v,t)] + \gamma T \frac{\partial^2}{\partial v^2}P(x,v,t)\\
J_{col}[v|P(x,v,t)]&=\int dv'[W_\epsilon(v|v')P(x,v',t)-W_\epsilon(v'|v)P(x,v,t)].
\end{align}
\end{subequations}
In Eq.~\eqref{model}, $U(x)$ is an (optional) external potential,
while $F_{nl}(v)$ represents the effect of non-linear dissipative
force, for which we assume that $v F_{nl}(v) \le 0$. A realistic
prototype of this force is Coulomb friction, which acts between
sliding rough surfaces and takes the form
\begin{equation}
F_{nl}(v)=-\Delta \sigma(v),
\end{equation}
where $\Delta$ is friction intensity and $\sigma(v)$ is the sign of $v$ (and $\sigma(v)=0$ when
$v=0$). 

Two ``bath'' terms, $J_{Lang}$ and $J_{col}$, are present for larger
generality: however -- in the examples discussed below -- they are
mutually exclusive, i.e. only one of the two is used~\footnote{We
  anticipate that in the ``diffusive'' limit of small gas particles,
  the collisional noise $J_{col}$ takes the form of a Langevin noise,
  equivalent to $J_{Lang}$ }. The Langevin term $J_{Lang}$ represents
the interaction with a heat-bath at temperature $T$
(Boltzmann's constant is put to $1$), with a thermalization time
$1/\gamma$. This term vanishes for a Gaussian steady state at
temperature $T$, i.e. $J_{Lang}[v|G_T(v)]=0$ with
$G_T(v)=\frac{1}{\sqrt{2 \pi T}}e^{-v^2/(2T)}$. The ``collisional''
bath term, $J_{col}$ takes the form of a Master Equation for jump
processes: in this term, $W_\epsilon(v|v')$ represents the rate for
the transition $v' \to v$ when the tracer is in contact with a very
large volume of a gas of hard-core particles of mass $\epsilon^2$, with the assumption that the
velocity distribution of the particles of the gas is not affected by
collisions with the tracer~\cite{vK61}. This occurs, for instance, when those
collisions are very rare with respect to collisions between two gas
particles, a condition which also implies Molecular Chaos for
tracer-gas collisions (we will always assume it). The particular form
of $W_\epsilon(v|v')$ depends on the kinematic of the gas-tracer
collision, e.g. on the gas density and the geometric shape of the
tracer. Different possibilities will be considered below, with
explicit examples of $W_\epsilon(v|v')$. In most of the cases we consider the gas of particles to be at
equilibrium at a temperature $T$. In the case of elastic
collisions, the rates $W_\epsilon(v|v')$ satisfy detailed balance with
respect to $G_{T}(v)$, which however does not imply that detailed
balance holds for the model, because of the presence of
$F_{nl}(v)$. In previous works it has been considered the more general case where the collisions between the tracer
and the gas particles can also be inelastic: however this is an {\em
  additional} mechanism of dissipation, not strictly necessary to get
a motor effect, introduced in order to describe granular
experiments~\cite{gnoli}. This mechanism is not discussed here.

We conclude the introduction of the general model, by mentioning that
a motor (or ``ratchet'') effect can be obtained in a steady state
only if detailed balance is broken and a spatial asymmetry is
present. As discussed in the examples below, the spatial asymmetry
~\footnote{A spatial asymmetry could also be present in the friction
  term, but we will not consider such a case here.} may be explicitly
present in the potential $U(x)$ or in the shape of the object. In the
last case, it appears encoded in the transition rates
$W_\epsilon(v'|v)$.

\subsection{Conditions to break detailed balance}

As announced, a motor effect requires the absence of symmetry under
the operation of time-reversal, which in our Markovian model is
equivalent to the breakdown of detailed balance condition. When the
non-linear friction is absent, i.e. $F_{nl}(v)=0$, the model satisfies
detailed balance, reaching a steady state with $P(x,v) \propto
e^{-U(x)/T}e^{-v^2/(2T)}$. In the absence of non-linear friction,
mechanisms to break detailed balance are the introduction of inelastic
collisions~\cite{costantini1} in the rates $W_\epsilon(v'|v)$ or unbalancing the
temperatures of the two baths defined by $J_{Lang}$ and
$J_{col}$. These mechanisms are not discussed in this paper, where
elastic collisions and baths at the same temperature are always
considered.


A point which is not much discussed in the literature, is the
following: non-linear friction ($F_{nl}(v) \neq 0$) does not break
detailed balance in a simple Langevin model, i.e. with $U(x)=0$ and
$W_\epsilon(v'|v)=0$. Indeed, in that case, the Fokker-Planck equation gets an
``equilibrium'' steady state with $P(x,v,t) \propto e^{-H(v)/(\gamma
  T)}$, with $H(v)=\gamma v^2/2-\int^v F_{nl}(v') dv'$.
See also the discussion in \cite{HG12}, and in \cite{DHG09,Po13}
for the case of multiplicative noise.

Cases which have been demonstrated to break detailed balance with
$F_{nl} \neq 0$ are: 1) in the presence of collisional noise,
$W_\epsilon \neq 0$ (even for elastic collisions)~\cite{gnoli,SGP13}; 2) in the presence
of a spatial potential, $U(x) \neq 0$~\cite{sarra}. We consider three possible
examples of the first case, Section~\ref{collisional}, and one example
of the second case, Section~\ref{sec_kramers}.

\subsection{Other models with non-linear frictions}

It is interesting to notice that in the literature many other models
and experiments featuring a ratchet-like effect have appeared, where
non-linear friction is an important ingredient. Some of these works
involve experimental observation of an average drift in the sliding
motion between vibrated surfaces, also of biological origin, in
several different
setups~\cite{mahadevan,daniel,eglin,buguin,fleishman,baule}.  

In all those works friction is counterbalanced by mechanisms for energy
injection which are non-thermal. In particular in~\cite{mahadevan} and
in~\cite{daniel} energy is injected by mechanical periodic vibrations
of the plate supporting the substrate; in~\cite{fleishman} the length
at rest of the springs connecting three massive blocks is periodically
modulated in time; in~\cite{eglin} a substrate is posed on a shear
polarized piezoelectric plate which is excited by a periodic electric
signal; in~\cite{buguin} a model with generic periodic acceleration is
considered; and finally in~\cite{baule} a model with random
acceleration in the form of a Poissonian shot noise is considered,
with explicit calculations performed for an exponential distribution
of the amplitude of the random kicks.

Already at a first look one realizes that the above mechanisms do not
closely correspond to thermal fluctuations. 
In our opinion none of the above mechanisms may mimic a thermal
bath. This can be understood from both a physical and a mathematical
point of view:

\begin{itemize}

\item physically, a thermal bath gives {\em and takes} energy to/from
  the system in such a way that - if other dissipations are switched
  off - the system remains at the same temperature of the bath: this
  is never verified in the model/experiments considered above. As a
  matter of fact all the above systems consist in a combination of two
  or three basic ingredients: (a) dry friction, (b) other dissipations
  (e.g. viscous friction, not present in all cases), (c) external
  energy injection. If all the dissipations (a) and (b) are removed,
  their energy will increase indefinitely; if only friction is removed
  and some other dissipation (b) is retained, the system will reach a
  balance of energy coming from (c) and going into (b); therefore
  there is a non-zero {\em current} of energy and the attained
  stationary state is clearly a non-equilibrium one;

\item mathematically, all those systems - when friction is removed -
  do not satisfy time-reversal symmetry; for instance the models
  considered in~\cite{buguin} and in~\cite{baule}, which are Markovian,
  do not satisfy detailed balance.

\end{itemize}

Summarizing, in all the above models/experiment the system is already
out-of-equilibrium, even without the presence of dry friction. The
model in Eq.~\eqref{model} is of a different nature: here
the energy dissipated by friction is balanced by a thermostatting
mechanism ($J_{Lang}$ and/or $J_{col}$) which is
a thermal bath precisely in both senses discussed above.

\section{Collisional noise}
\label{collisional}

In this Section we review some examples of Eq.~\eqref{model} with only
collisional noise, i.e. $J_{Lang}=0$ and $J_{col} \neq 0$, and no need
for the external potential, $U(x)=0$. The first two examples are
idealised models for a translational piston in contact, through
elastic collisions, with a gas of particles: the first one takes into
account an over-simplified collision rule and an {\em asymmetric}
distribution of the gas particles's velocities, with zero average
velocity; the second example treats hard-core collisions with a gas at
equilibrium, with a spatial asymmetry introduced by considering
different masses for the particles hitting the piston from the left or
from the right. The third example concerns the dynamics of a {\em
  rotator} colliding with particles of a gas at equilibrium: spatial
symmetry is broken by considering an asymmetric shape of the rotator.

The explicit expression of the transition rates appearing in
$J_{col}$, see Eq.~\eqref{model}, is different for each particular
case. An almost general expression is presented here to explain the
basic idea:
\begin{equation} \label{colrate}
  W_{\epsilon}(v'|v)=\rho S \int_S \frac{ds}{S} \int d \bu \phi(\bu) (v \hat{x}-\bu) \cdot \hat{n} \Theta[(v \hat{x}-\bu)\cdot \hat{n}] \delta[v'-v-\delta v(v,\bu,\epsilon,s)],
\end{equation}
where $\rho$ is the gas density, $\bu$ represents the velocity of a gas particle, $\phi(\bu)$ its
pdf, $\Theta$ is the Heaviside function and $\delta
v(v,\bu,\epsilon,s)$ is the change of velocity in a collision between
the intruder at velocity $v$, the gas particle of mass $\epsilon^2$ at velocity $\bu$, and a position of impact on the intruder
surface parametrized by the curvilinear ascissa $s$, where the normal
going out from the surface has direction $\hat{n}$. We assume to be
in two dimensions and that the motor has total impact surface $S$. For this particular example
(similar to the one presented in Sec. III.C), it is restricted to move
along the $\hat{x}$ direction. The two main physical assumptions
justifying Eq.~\eqref{colrate} are Molecular Chaos, typically
justified by diluteness of the gas, and {\em independence} of the gas
from the state of the intruder, which allows one to keep $\phi(\bu)$
as a constant parameter of the problem. The term $(v \hat{x}-\bu)
\cdot \hat{n}$ represents a hard core interaction potential (but one
can make more general choices, of course). In most of the calculation
below, $\phi(\bu)$ is assumed to be Gaussian.

Before entering the discussion of the different examples, we recall
that in the limit of very light gas particles the effect of the
collisional noise tends to become equivalent to the effect of a
Langevin bath.

\subsection{White noise limit}

Assuming that the surrounding gas has an equilibrium Gaussian velocity
distribution at temperature $T$, in the limit of small mass of the gas
particles $\epsilon \to 0$ (we recall that we set to $1$ the mass of
the intruder), one can simplify the integro-differential
Equation~\eqref{model}. Performing a Van Kampen expansion \cite{vK61}
up to the second order, the master equation contribution $J_{col}$
reduces to the the sum of a linear viscous friction term and an
uncorrelated white noise
\begin{eqnarray} \label{difflimit}
&&\int dv'[W_\epsilon(v|v')P(x,v',t)-W_\epsilon(v'|v)P(x,v,t)]\rightarrow  \nonumber \\
&&-\frac{\partial}{\partial v}[(-\gamma_g v + F_g) P(x,v,t)]+\frac{\partial^2}{\partial v^2}[\gamma_g T P(x,v,t)],
\end{eqnarray}
where $\gamma_g$ and $F_g$ depend on the particular form of the original transition rates $W_\epsilon(v|v')$.

Putting Eq.~\eqref{difflimit} in Eq.~\eqref{model} (setting for simplicity $\gamma=0$), one obtains the following equation for the drift at stationarity
\begin{equation}
\langle v\rangle=\frac{F_g}{\gamma_g}+\frac{\langle F_{nl}(v)\rangle}{\gamma_g}.
\label{2.6}
\end{equation}
In the many examples discussed in the literature, as well as on the basis of general arguments, it is 
observed that the constant force $F_g$ takes a simple form of the kind
\begin{equation}
F_g = \mathcal{A} (T_r-T)
\label{2.7}
\end{equation}
where $T_r=\langle v^2 \rangle$ is the ``tracer temperature'', and $\mathcal{A}$ is a coefficient denoting the spatial asymmetry in
the system. This general form will be reproduced in all the examples
discussed below.

Eq.~\eqref{2.7} shows that, in the absence of non-linear friction, a
ratchet effect can be present if the rates do not satisfy detailed
balance (e.g. when collisions are inelastic), so that the tracer
temperature $T_r$ is different from that of the external bath
$T$. However, we stress that a non-zero drift can also be obtained
when the rates do satisfy detailed balance (e.g. for elastic
collisions): the presence of non-linear friction reduces the average
ratchet energy so that $T_r < T$.

\subsection{Flat collision rule}

Consider the following rates, which only depend on the final state:
\begin{equation}
W_\epsilon(v'|v)=\frac{f(v')}{\tau_c}.
\label{1.3}
\end{equation}
Here $f(v)$ is the probability density function for the {\em
  post-collisional} velocity and $\tau_c$ is the mean time between two
collisions.  We are therefore considering a collisional process where
at each collision the state is completely independent from the
previous one.  This assumption, which over-simplifies the interaction
between the ratchet and the environment, allows us to find out exact
results about the motion of the system in the stationary state.  We
want to stress that this kind of process represents a legitimate heat
bath: indeed, in the absence of all other dissipative terms (that is
Eq.~\eqref{model} with only $J_{col} \neq 0$), and assuming for simplicity $U(x) \equiv 0$, a steady state is
reached with steady pdf $P(x,v)=f(v)$, such that detailed balance is
trivially satisfied. Even in this case, the model we are considering
is different from other proposed models with ``simple'' collisions
rules, e.g. from~\cite{baule}: in our case the instantaneous change of
velocity is $v'-v$, which is correlated with $v$, while
in~\cite{baule} the instantaneous change of velocity is a Poissonian
process totally independent from $v$.

We consider the Coulomb friction law for $F_{nl}(v)$ so that between two collisions, the ratchet follows the motion
equation $\dot{v} = - \Delta \sigma(v)$.
The parameters of the system such as temperature and spatial asymmetry are 
fully contained in the pdf $f(v)$, and the breaking of the time-reversal
symmetry is ensured by the presence of friction.\\

\begin{figure}
\includegraphics[width=8cm,clip=true]{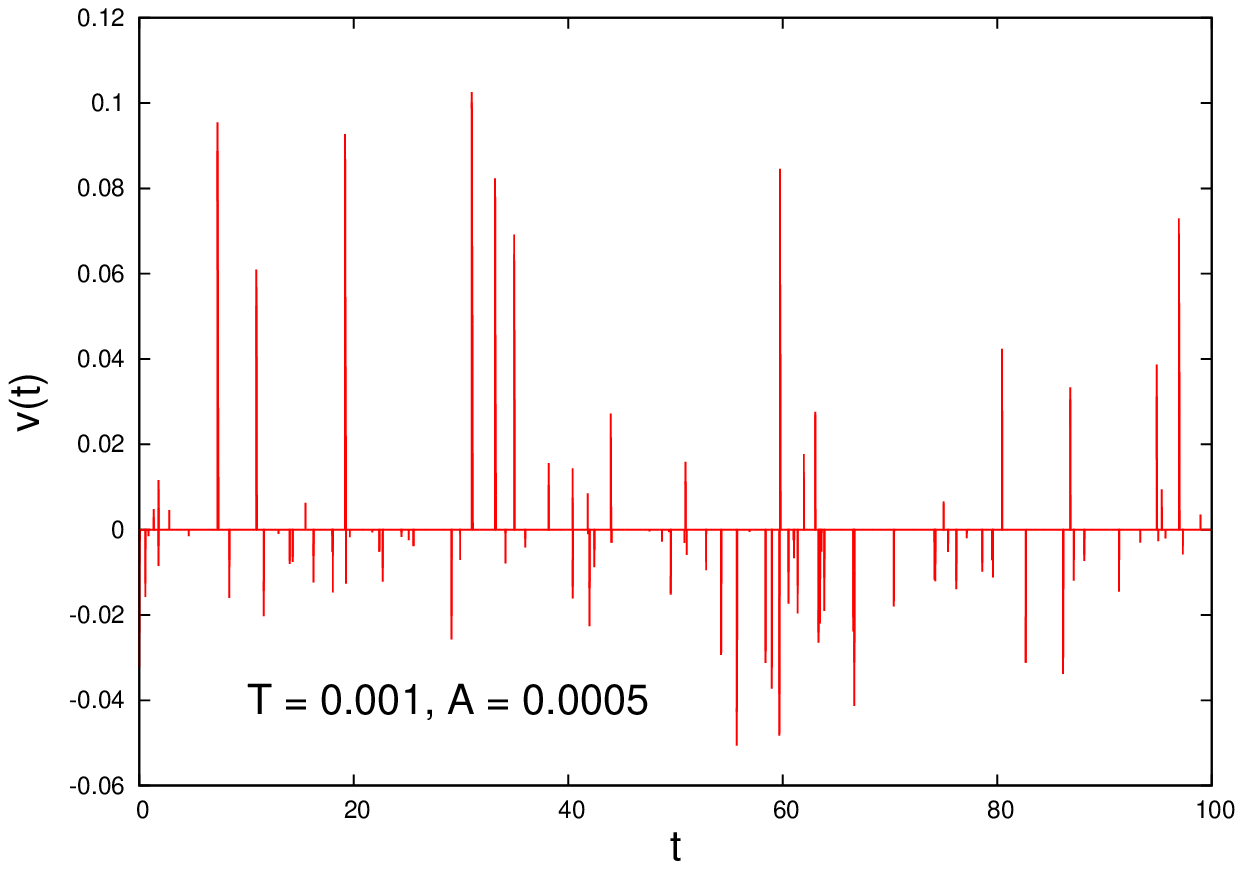}
\includegraphics[width=8cm,clip=true]{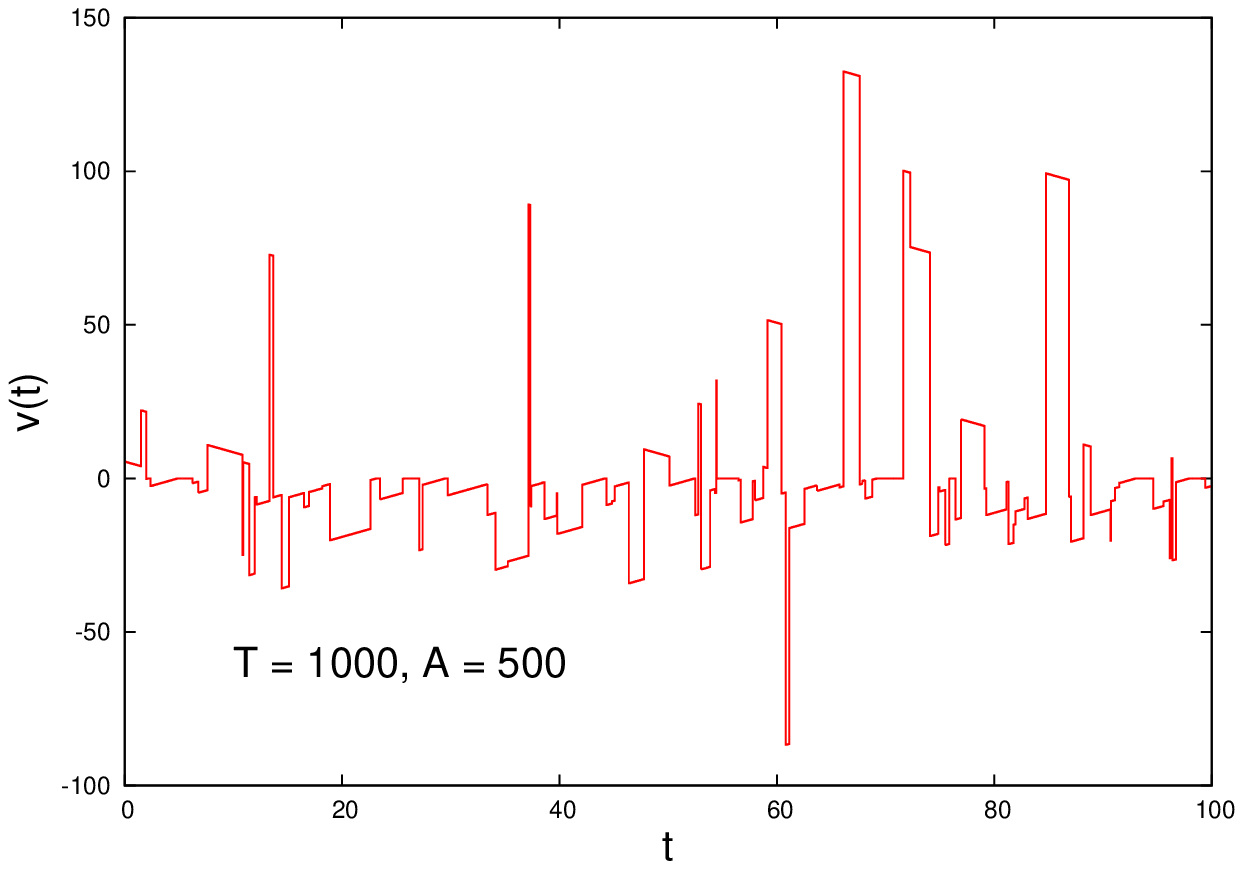}
\caption{Left: Velocity vs time for the ratchet, when asymmetry is present and $\tau_c /
\tau_{\Delta} \simeq 10^3$ (rare collisions limit). Simulations of the dynamics are performed 
using a $f(v)$ exponential, with different decay rates for positive or negative $v$. 
Right: same plot, with $\tau_c /\tau_{\Delta} \simeq 10^{-3}$ (frequent collisions limit). 
See caption of Fig.~\ref{velAM} for the definitions of $A$ and $T$.} 
\label{trajAM}
\end{figure}

Supposing the system to be ergodic, we can compute the stationary averages of 
the dynamical variables of the ratchet by performing their time average over long trajectories. 
The calculation of time averages is possible noticing that between two collisions 
the ratchet velocity is $v(t) = v_0 - \sigma(v_0) \Delta t$, where $v_0$ is the velocity after 
that collision; this motion introduces another time scale, the mean stopping time 
of the ratchet $\tau_{\Delta} = \langle \vert v \vert \rangle_f / \Delta$ (where we denote with 
$\langle \cdot \rangle_f$ the averages over the pdf $f(v)$). When $\tau_c \gg \tau_{\Delta}$, 
the ratchet generally stops before a new collision 
(rare collisions limit); viceversa, when $\tau_c \ll \tau_{\Delta}$ the ratchet is almost 
always in motion (frequent collisions limit). Simulated trajectories for both cases are shown in 
Fig.~(\ref{trajAM}). Performing the time average, we find the average velocity 
of the ratchet
\begin{equation} \label{v1AM}
\langle v \rangle = \langle v \rangle_f - \bar{v} \langle \sigma(v) ( 1 - e^{-|v|/\bar{v}} ) \rangle_f ,
\end{equation}
where $\bar{v} = \Delta \tau_c$ is a characteristic velocity. 
Through the characteristic function $q(k) = \langle e^{ikv} \rangle$, it is possible to compute the 
stationary probability density of the velocity that is
\begin{equation} \label{PAM}
P(v) = \int^{+\infty}_{-\infty} dv' f(v') \left\lbrace e^{-|v'|/\bar{v}} \delta(v) + 
\frac{\sigma(v')}{\bar{v}} e^{(v-v')\sigma(v')/\bar{v}} \left[ \Theta(v' - v) - \Theta(-v) \right] 
\right\rbrace
\end{equation}
From Eq.~(\ref{v1AM}), we notice that the second term in the rhs yields a net drift also if $\langle v \rangle_f 
= 0$: this is the ratchet effect we are looking for. To observe it, it is necessary that 
$\Delta \neq 0$ (breaking of time reversal symmetry) and $f(v) \neq f(-v)$ (breaking of spatial 
symmetry). In the rare and the frequent collisions limits, if $\langle v \rangle_f = 0$ we find 
respectively that
\begin{subequations} \label{v_limAM}
\begin{align}
\tau_c \gg \tau_{\Delta} \quad & \Rightarrow \quad \langle v \rangle 
\approx \frac{1}{2 \bar{v}}\langle \sigma(v) \, v^2 \rangle_f \\
\tau_c \ll \tau_{\Delta} \quad & \Rightarrow \quad \langle v \rangle 
\approx - \, \bar{v} \, \langle \sigma(v) \rangle_f
\end{align}
\end{subequations}
so, depending on $f(v)$, the sign of the velocity can change by changing the parameters 
of the system. Furthermore, from Eq.~(\ref{PAM}), we point out the presence of a delta function 
into $P(v)$: its weight represents the finite time that the ratchet spends in $v=0$. Generally, 
the stationary pdf of the velocity for ratchet with Coulomb friction can be written as~(\cite{talbot2})
\begin{equation} \label{PgenAM}
P(v) = \gamma_0 \delta(v) + \gamma_R P_R (v) ,
\end{equation}
where $P_R (v)$ is a regular function of $v$. In our model, we obtain $\gamma_0 = \left\langle	
e^{-v/\bar{v}} \right\rangle_f $, that goes to 1 in the rare collisions limit, and to 0 in 
the frequent collisions limit.\\

\begin{figure}
\includegraphics[width=8cm,clip=true]{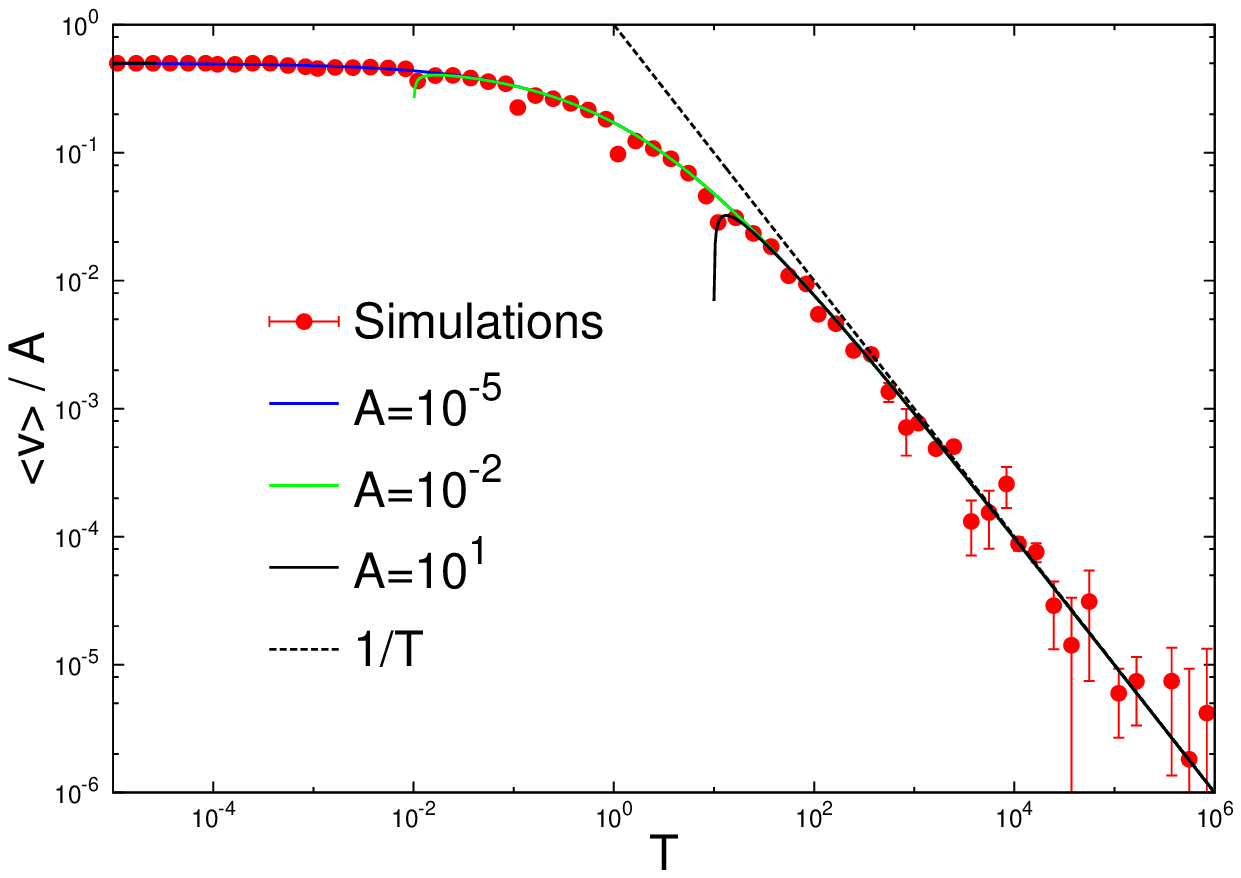}
\includegraphics[width=8cm,clip=true]{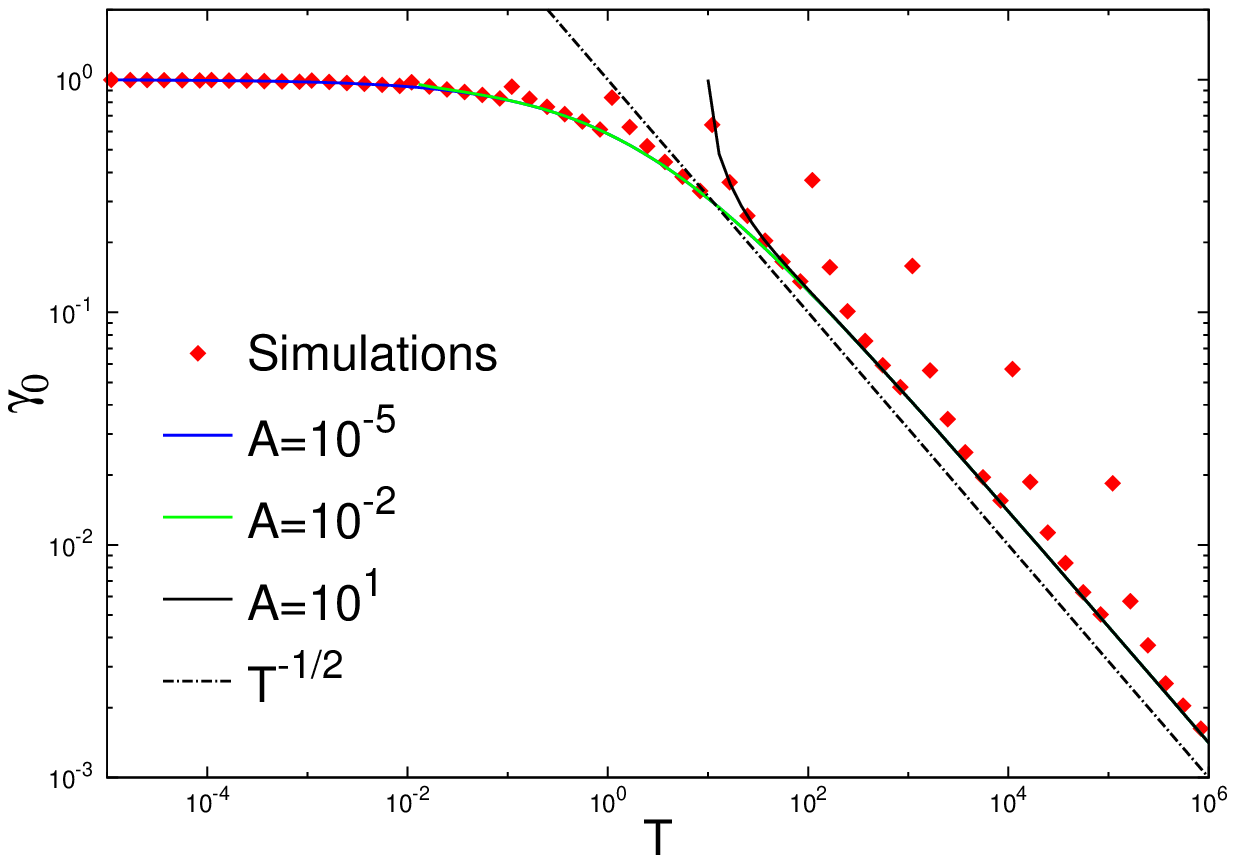}
\caption{Left: Numerical simulations and analytical results for the
  average velocity of the ratchet, Eq.(\ref{v1AM}), rescaled by $A =
  \langle \sigma(v) v^2 \rangle_f$, which characterizes the asymmetry
  of the system ($A=0$ for even $f(v)$). The temperatures $T$ are
  given by $T = \langle v^2 \rangle_f$ (we are also assuming $\langle
  v \rangle_f=0$, and we set $\tau_c = \Delta = \bar{v} = 1$, then
  $\tau_{\Delta} \simeq \sqrt{T}$.  $f(v)$ is chosen as in
  Fig.~(\ref{trajAM}). The dashed line represents the $T^{-1}$ asymptotic behavior 
  predicted in this case for large values of $T$. Right: Numerical simulations and analytical
  results for $\gamma_0$. Same parameters as in left plot. The dotted
  line represents the $T^{-1/2}$ decay predicted in this case for large $T$.}
\label{velAM} 
\end{figure}

We simulate numerically the dynamics of the system, using a $f(v)$
exponential with different decay rates for $v$ positive or negative,
finding perfect agreement between simulations and theory; some results
are shown in Fig.~(\ref{velAM}), putting in evidence the asymptotic
decays for $\langle v \rangle$ and $\gamma_0$ at large $T$.

\subsection{The asymmetric Rayleigh piston} 

\begin{figure}[t!]
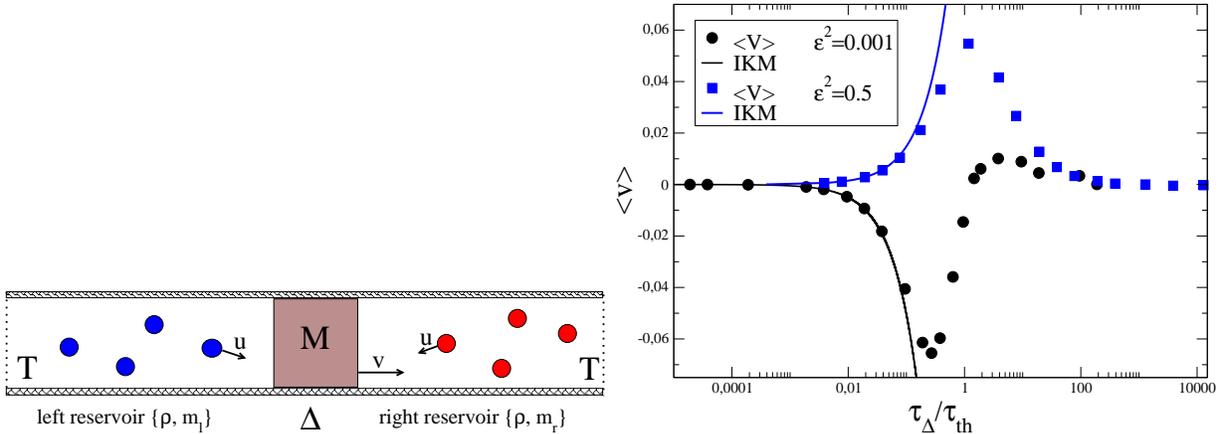

\includegraphics[width=8cm,clip=true]{fig_pist.eps}
\includegraphics[width=8cm,clip=true]{fig_last.eps}
\caption{Left: a sketch of the asymmetric Rayleigh piston model (in
  this paper the piston mass $M$ is set to $1$). Right: Numerical
  simulations of the Rayleigh piston, with $T=10$, $\rho=0.5$,
  $\alpha=2$, showing $\langle v\rangle$ as a function of
  $\tau_\Delta/\tau_{th}$, for $\eps^2=0.01$ (black dots) and
  $\eps^2=0.5$ (blue squares). The continuous curves show the
  analytical predictions of the IKM, Eq.~(\ref{drift}).}
\label{piston} 
\end{figure}

We study here the dynamics of an asymmetric Rayleigh
piston, introduced in~\cite{macdonald}, under the action of Coulomb friction \cite{SGP13,tomo}. In this
model a piston of mass $1$, constrained to move
along a given direction, interacts via elastic collisions with two gases of particles,
which are placed on both sides of the piston (see Fig.~\ref{piston}).
The particles of the two gases have different masses but are kept at the same temperature $T$. 
Therefore, in the absence of nonlinear friction, the whole system is at equilibrium
at temperature $T$ and the spatial asymmetry introduced by the different masses cannot
produce any rectification of fluctuations. On the contrary, when the motion of the piston
is also subjected to Coulomb friction, 
dissipative effects intervene and a motor effect is observed.

The dynamics of the piston is described by Eq.(\ref{model}), neglecting the spatial dependence ($U(x) \equiv 0$), 
with $J_{Lang}[v|P(v,t)]=0$ and $F_{nl}=-\Delta\sigma(v)$.
Denoting by $m_l=\eps^2$ and $m_r=\alpha \eps^2 $ the masses of particles at left and at right of the piston
respectively, and by $p_r(u)$ and $p_l(u)$ their velocity distributions, 
where $u$ is the velocity of gas particles, the asymmetric transition rates appearing
in the collision term $J_{col}[v|P(v,t)]$ are~\cite{macdonald}
\begin{eqnarray}
W_\epsilon(v'|v) &=& \left(\frac{1+\eps^2}{2\eps^2}\right)^2(v'-v) \nonumber  \\
&\times& p_l\left(\frac{1+\eps^2}{2\eps^2}v'-\frac{1-\eps^2}{2\eps^2}v\right),  \nonumber \\
W_\epsilon(v'|v) &=& \left(\frac{1+\alpha\eps^2}{2\alpha\eps^2}\right)^2(v-v') \nonumber \\
&\times& p_r\left(\frac{1+\alpha\eps^2}{2\alpha\eps^2}v'-\frac{1-\alpha\eps^2}{2\alpha\eps^2}v\right).  
\label{ratep}
\end{eqnarray}
Notice that these transition rates satisfy DB with respect to the Gaussian
distribution $P_0(v)=(2\pi T)^{-1/2}\exp(-v^2/2T)$ even when $\alpha \neq 1$~\cite{macdonald}.  

In this model it is important to point out two time scales:
$\tau_\Delta=v^*/\Delta=\sqrt{T}/\Delta$, which is the stopping time
due to friction, and $\tau_{th}\simeq
\sqrt{\pi/(2T)}/[2\rho(\sqrt{\eps^2}+\sqrt{\alpha\eps^2})]$, which is the
thermalization time of the piston with the gas in the absence of
friction (proportional to the mean collision time), where $\rho$ is
the gas density (equal on both sides of the piston).

An analytical expression for the
average drift can be obtained in the limit of rare
collisions, namely when $\tau_{th}\gg\tau_\Delta$. In this case,
assuming that every collision occurs when the piston is at rest, the
average velocity can be computed using the Independent Kick
Model (IKM) introduced in~\cite{talbot1,talbot2}.  For our model this yields
\begin{equation}
\langle v\rangle = \left(\int du |u|p_r(u)+\int du |u|p_l(u)\right)\int_0^\tau v(t)dt,
\end{equation}
where $v(t)=v_0-\Delta\sigma(v_0) t$, $\tau=|v_0|/\Delta$ and $v_0$ is
the velocity after a collision: $v_0=v^+$ if $u>0$, and $v_0=v^-$ if
$u<0$, where $v^+=\frac{2u}{1+1/(\alpha\eps^2)}$ and $v^-=\frac{2u}{1+1/\eps^2}$.
Using these expressions, and considering a gaussian distribution for $p_{r/l}(u)$ 
with variance $T/m_{r/l}$, one obtains
\begin{equation}\label{drift}
\langle v\rangle=\frac{2\rho}{\Delta}\sqrt{\frac{2T^3}{\pi}}
\left[\frac{\sqrt{\eps^2}}{(\eps^2+1)^2}
-\frac{\sqrt{\alpha\eps^2}}{(\alpha\eps^2+1)^2}\right].
\end{equation}
Notice that the average velocity is finite when the asymmetry in the
system is present (i.e. $\alpha \ne 1$).  Notice also that in the
limit $\eps\to 0$ the drift vanishes.  In Fig.~\ref{piston}, right
panel, the analytical prediction~(\ref{drift}) of the IKM (black
lines) is shown to be in perfect agreement with the numerical results
(see \cite{SGP13} for details) in the rare collision regime.
Fig.~\ref{piston}, right panel, also shows $\langle v\rangle$ (black
dots for $\eps^2=0.01$ and blue squares for $\epsilon^2=0.5$, with
$\alpha=2$) for a large range of values of the ratio
$\tau_\Delta/\tau_{th}$, which is varied by changing $\Delta$, with
the other parameters fixed (see caption). A net drift is observed, as
expected. It is interesting to notice that thermodynamic (bulk)
pressures are equal in the two reservoirs and therefore the motion of
the piston is due to the fact that the average exchanged momentum
between gas and piston is not equal to bulk pressure. A recent theory
about non-equilibrium momentum deficit~\cite{FKS12} is difficult to be
applied in this particular case, as noticed recently by other
authors~\cite{tomo}.  Another interesting feature which appears in
simulations, but is hardly explained by theory, is the presence of
extremal points (even more than one) and current inversion points. A
similar behavior is encountered in other models discussed here.

\subsection{Rotator with an asymmetric shape}

\begin{figure}[t!]
\includegraphics[width=6cm,clip=true]{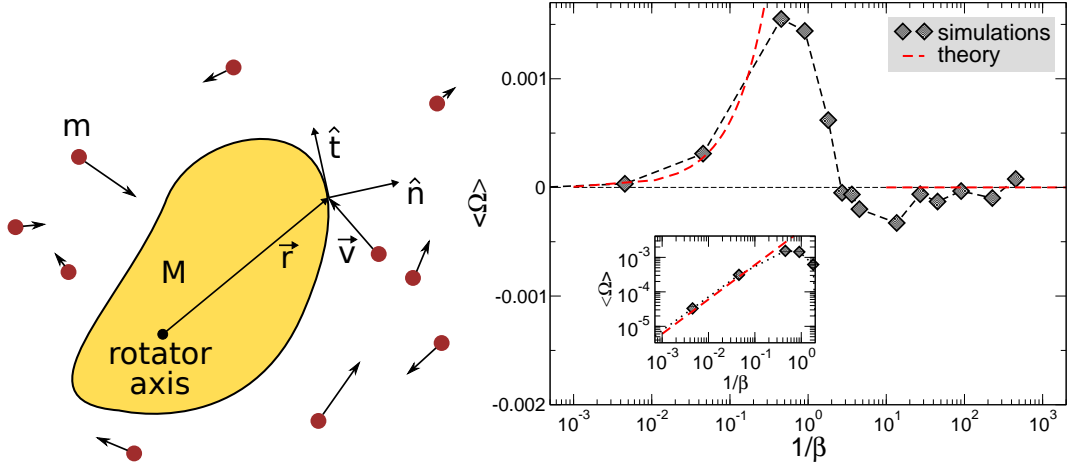}
\includegraphics[width=8cm,clip=true]{numerical.eps}
\caption{Left: sketch of the model. Right: results of numerical simulations for the rotator model with asymmetric shape, for different values of $\beta^{-1}$.}
\label{rotator} 
\end{figure}

In this subsection, the tracer driven by the
Brownian motor effect is a {\em rotator}, and for this reason we
replace $x,v$ by $\theta,\omega$, i.e. angular displacement and
angular velocity respectively.  The rotator has momentum of inertia
$I$, mass $1$ and radius of inertia $\ri = \sqrt{I}$. 
It is constituted by the set of material points with
cartesian coordinates $\{x,y,z\}$ with $z\in[0,h]$ (where $h$ is its
height) and $\sqrt{x^2+y^2}<r(s)$ for each $s\in[0,S]$ where $s$ is
the curvilinear abscissa, $r(s)$ is the curve delimiting a section of
the solid in the $\hat{xy}$ plane, and $S$ is the perimeter of the
section.  The rotator changes its angular velocity for two reason: 1)
because of dry friction $F_{nl}=-\Delta \sigma(\omega)$, with $\Delta$
the frictional torque rescaled by momentum of inertia, and 2) for the
effect of elastic collisions with a dilute gas of particles at
equilibrium at temperature $T=\eps^2 v_0^2$. The gas surrounding the rotator
has volume number density $n$ and $\rho=n h$ is its two-dimensional
projection, which is the only one which matters in the problem. Note
that $\rho S \equiv n \Sigma$ with $\Sigma$ the total surface of the
rotator parallel to the rotation axis.  We finally assume that no
external potential and no heat bath are present,
i.e. Eq.~\eqref{model} (with the replacements $x \to \theta$, $v \to
\omega$) holds with $U\equiv 0$ and $\gamma=0$. A sketch of the model with used symbols is shown in Fig.~\ref{rotator}, left panel.

The effect of the elastic collisions with the equilibrium gas is to
change $\omega$ into $\omega'$ and that of the colliding particle from
$\bu$ to $\bu'$, following the rule
\begin{subequations} \label{col_rule}
\begin{align} 
\omega'&=\omega+2\frac{(\bv- \bu) \cdot \hatn}{\ri}\frac{g\eps^2
}{1+\eps^2 g^2},\\
\bu' &= \bu + 2\frac{(\bv- \bu) \cdot \hatn}{1+ \eps^2 g^2}\hatn
\end{align}
\end{subequations}
where $\bv = \omega \hat{z} \times {\mathbf r}$ is the linear velocity of the rotator at the point of
impact ${\mathbf r}$, $\hat{n}$ is the unit vector perpendicular to
the surface at that point, and finally $g=\frac{{\mathbf r}\cdot
  \hat{t}}{\ri}$ with $\hat{t}=\hat{z} \times \hat{n}$ which is the
unit vector tangent to the surface at the point of impact.
Equations~\eqref{col_rule} guarantee that total angular momentum and
total kinetic energy are conserved and that relative velocity
projected on the collision unit vector is reflected.  A few relations
in cartesian coordinates may be useful: $\bv=(-\omega r_y, \omega
r_x)$ and $\hatt=(-n_y,n_x)$. It is also useful to realize that $\bv
\cdot \hatn=-\omega \ri g$, and to introduce the ``equilibrium'' angular velocity $\omega_0=\frac{\eps v_0}{\ri}$ and the rescaled velocity $\Omega=\frac{\omega}{\omega_0}$. 

With the above collision rule, the transition rates $W_\epsilon$ take the explicit form
\begin{multline}
W(\omega'|\omega)=\frac{\rho S \ri^2}{8 \sqrt{2\pi} \eps^2 v_0}\int
\frac{ds}{S} |\omega'-\omega|\frac{(1+\eps^2g^2)^2}{\eps^2g^2} \\
\times\Theta\left[\frac{\omega'-\omega}{g}\right]\exp\left[-\frac{\ri^2}{2\eps^2
v_0^2} \left(\omega \eps g+\frac{(\omega'-\omega)(1+\eps^2g^2)}{2\eps
g}\right)^2\right].
\end{multline}.

Following the same lines of the two previous sections, it is
useful to  realize that two time-scales are relevant in the system: 1) the mean stopping
time due to environmental dissipation, which is dominated by dry
friction (being almost always $\gat |\omega| < \dat$),
$\tau_\Delta=\frac{\langle |\omega|\rangle_{pc}}{\dat} \sim \frac{\eps
  v_0}{R_I \dat}$, where $\langle \cdot \rangle_{pc}$ denotes a
post-collisional average; 2) the mean free time between two collisions
$\tau_c \sim \frac{1}{n \Sigma v_0}$. This implies the existence of a main control parameter
\begin{equation}
\beta^{-1}=\frac{\eps n\Sigma v_0^2}{\sqrt{2} \pi \ri \dat}  \approx
\frac{\tau_\Delta}{\tau_c}
\end{equation}
which is an estimate of the ratio of those two time-scales, as
verified in simulations.

When the mass of the rotator is large, $\eps \ll 1$, and $\beta^{-1}
\gg 1$, friction becomes negligible and the collisional noise becomes
white noise, so that the average drift tends to zero. A finite drift
could be achieved in the case of inelastic collisions, which has been
considered elsewhere~\cite{gnoli,GSPP13}.

In the opposite limit $\beta^{-1}\ll 1$, an independent kicks approximation leads to the formula for
the rescaled average velocity of the ratchet
\begin{subequations} \label{rcl}
\begin{align} \label{rcla}
\langle \Omega \rangle  &=  \sqrt{\pi}4\beta^{-1} \eps^2 \mathcal{A}_{RCL}\\
\mathcal{A}_{RCL} &= \left\langle
\frac{ \sigma(g)g^2}{(1+\eps^2 g^2)^2} \right\rangle_{surf},
\end{align}
\end{subequations}
where $\mathcal{A}_{RCL}=0$ for symmetric shapes of
the rotator. Above we have used the shorthand notation for the uniform average
along the perimeter (denoted as ``surface'') of a horizontal section
of the rotator $\langle  \rangle_{surf} = \int_{surf} \frac{ds}{S}$.
Note that the limit
$\Delta \to 0$ is singular in formula~\eqref{rcla}, since in the
absence of dissipation between collisions the stopping time becomes
infinite, $\tau_\Delta \to \infty$, and the assumption of ``rare collisions'', $\beta^{-1}\ll 1$, breaks down.
The magnitude of the drift is predicted to increase
with $\beta^{-1}$: this corresponds to $|\langle \omega \rangle| \sim
v_0^3$.

In Figure~\ref{rotator}, right panel, we have shown the results of
numerical simulations for the rotator model with a shape identical to
the one used in recent granular experiments, see Ref.~\cite{gnoli} for
details. Both predictions for large and small values of $\beta^{-1}$
are well reproduced. As expected by interpolating the two predictions,
the average drift velocity $\langle \Omega \rangle$ goes through a
maximum. A still unexplained feature, common also to the Rayleigh piston case discussed previously, is the presence of a current
inversion and a second extremal point (a minimum), before going
asymptotically to zero at large $\beta^{-1}$.

\section{Kramers equation with nonlinear friction}
\label{sec_kramers}

In this section we consider the effect of an asymmetric spatial potential $U(x)$,
coupled to nonequilibrium conditions induced by the presence of
nonlinear velocity-dependent forces $F_{nl}$~\cite{sarra}. In particular, 
we consider a "generalized" Klein-Kramers equation for the motion of a
particle of mass $m=1$, with position $x$ and velocity $v$, subjected to thermal fluctuations,
\begin{eqnarray}\label{kramers}
\dot{x}(t) &=& v(t) \nonumber \\
\dot{v}(t)&=&-\gamma v(t) + F_{nl}[v(t)]-U'[x(t)] +\eta(t), 
\end{eqnarray}
where $\eta(t)$ is a white noise, with $\langle \eta(t)\rangle=0$ and
$\langle \eta(t)\eta(t')\rangle=2\gamma T\delta(t-t')$, $\gamma$ and
$T$ being two parameters and $\delta(t)$ the Dirac's delta. 
The corresponding Fokker-Planck equation for this model 
is given by Eq. (\ref{model}), with $J_{col}=0$.

We consider here the nonlinear force in the form of Coulomb friction, namely
\begin{equation}\label{coulomb}
F_{nl}[v(t)]=-\Delta\sigma[v(t)]. 
\end{equation}
Without external potential, model~(\ref{kramers}) with friction~(\ref{coulomb}) 
has been studied for instance in~\cite{dGen05,H05,DCdG05,PF07,BTC11}.

We also consider a model for active Brownian
particles~\cite{SET98}, inspired by the Rayleigh-Helmholtz model for
sustained sound waves~\cite{rayleigh}, where
\begin{equation}\label{active}
F_{nl}[v(t)]=\gamma_1 v(t)-\gamma_2 v^3(t),
\end{equation}
with $\gamma_1$ and $\gamma_2$ positive constants, and $\gamma=0$ in Eq.~(\ref{kramers}).
The motion of the particle is
accelerated for small $v$ and is damped for high $v$. This model
represents the internal energy conversion of the active
particles. 

The asymmetric ratchet potential is the one usually studied in the literature
of Brownian motors~\cite{BHK94}
\begin{equation}\label{ratchet_pot}
U(x)=\sin(x)+\mu\sin(2x), 
\end{equation}
where $\mu$ is an asymmetry parameter.
In the case of frictional force~(\ref{active}) the effect of an
asymmetric potential has been investigated in~\cite{STE00,FEGN08}. 

\begin{figure}[t!]
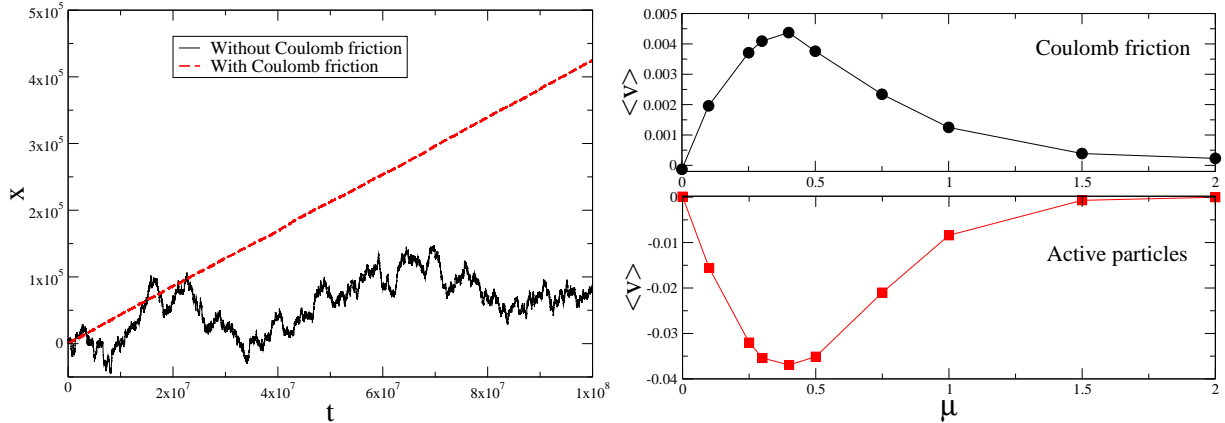

\includegraphics[width=8cm,clip=true]{traj.eps}
\includegraphics[width=8cm,clip=true]{drift_mu.eps}
\caption{Left: Evolution in time of the position $x(t)$ for the
  model in Eq.~(\ref{kramers}) without Coulomb friction (black
  continuous lines), with parameters $\gamma=0.05$, $\gamma T=0.5$,
  and $\mu=0.4$, and with Coulomb friction (red dashed line), with
  same parameters and $\Delta=1$. Right: Average drift for the model with Coulomb
  friction~(\ref{coulomb}) (top panel) with parameters $\Delta=1$,
  $\gamma=0.05$, $\gamma T=0.5$, and for the model for active
  particles~(\ref{active}) (bottom panel) with parameters
  $\gamma_1=\gamma_2=1$, and $\gamma T=0.5$, as a function of the
  parameter $\mu$.}
\label{traj} 
\end{figure}

We have performed numerical simulations of the model (\ref{kramers}),
which are reported in Fig.~\ref{traj}, left panel. Here we show the position of the
Brownian particle in time, in the absence
(continuous black line) and in the presence of Coulomb friction
(dashed red line). Notice the strong rectification phenomenon
occurring in the nonequilibrium case, namely when Coulomb
friction is present. 

In Fig.~\ref{traj}, right panel, we report the behavior of the system at varying
the parameter $\mu$ of the asymettric potential. In the top panel we
show the average velocity of the particle described by
Eq.~(\ref{kramers}) with Coulomb friction~(\ref{coulomb}) and in the
lower panel, we show the results of numerical simulations of the model
for active particles described by Eq.~(\ref{active}). As expected, for
$\mu=0$, the ratchet effect vanishes in both models, because the
potential is spatially symmetric in that case. By increasing the value
of $\mu$ a non-monotonic behavior is observed. The decreasing of the
ratchet effect for large values of $\mu$ is probably due to the fact
that the potential develops more than one minimum. This causes an
overall slowing down of the dynamics and, therefore, of the average
velocity.

\section{Conclusions}

We have reviewed a few models of Markovian Brownian motors: the common
ingredient is non-linear friction as the only mechanism for energy
dissipation. All other features, such as an external potential as well
as thermostats at the same temperature, are of ``equilibrium''
nature. Our focus is on a particular choice of non-linear friction,
which is found in many macroscopic experiments, that is Coulomb
friction between dry surfaces. The interplay between non-linear
friction and the other (deterministic and stochastic) forces acting on
the motor is subtle and not always generates a motor effect. We have
shown two possible routes toward rectification: they go through the
presence of thermal non-white noise (e.g. collisions with a gas at
equilibrium) or through the introduction of spatial inhomogeneity
(e.g. an external potential). What is common between these two
mechanisms is that they grant to the system a larger set of possible
trajectories (with respect to white noise in homogeneous space). In
this larger set, {\em cyclic} trajectories are accessible which
guarantee probability currents and (in the presence of a spatial
asymmetry) a motor effect. Apart from the simplest model discussed in
section III.B, full analytical treatment of these models is missing,
and expressions for the drift are known only in particular
limits. Far from these limits, numerical simulations suggest a rich and complex behavior,
with extremal and inversion points in the drift as a function of the
models' parameters, which are still lacking a satisfying explanation.

\begin{acknowledgments}
The authors acknowledge useful discussions with H. Hayakawa and T. Sano. The work of AS and AP is supported by the ``Granular-Chaos'' project, funded by the Italian MIUR under the FIRB-IDEAS grant number RBID08Z9JE. 
\end{acknowledgments}

\bibliography{fluct.bib}

\end{document}